# COMPUTER AIDED PLANNING AND NAVIGATION FOR ORBITO-ZYGOMATIC RECONSTRUCTION

CH. MARECAUX*,**, M. CHABANAS*,
Y. PAYAN*, F. BOUTAULT***

\* TIMC/GMCAO Laboratory, In3S - Université Joseph Fourier, Faculté de Médecine - 38706 La Tronche cedex.
\*\* Clinic Cours Dillon – 1, rue Peyrolade - 31300 Toulouse.
\*\*\* Department of maxillofacial and facial plastic surgery, Hopital Purpan, TSA 40031 - 31059 Toulouse cedex 9.

3D Navigation Systems (3DNS) were introduced in the late 80s in neurosurgery and no later described in maxillofacial surgery. Concomitant development of computed imaging techniques offered exciting horizons for surgical planning and simulation and principles of Computer Aided Surgery (CAS) could be formulated for a consistent use in Cranio-maxillofacial surgery. To date it has been widely developed and increasing fields of indications were reported. However, no 3DNS specifically designed for maxillofacial surgery or one pathology dedicated planning software platform are yet commercially available.

We suggest a full protocol of CAS as previously recommended in literature addressing the challenging task of primary or secondary reconstruction of OZMC dislocation. First, on a specifically developed planning software, the best zygoma reduction and orbital boundaries reconstruction to achieve skeletal symmetry are determined. This treatment plan is then transferred to the 3DNS within the operating room. After patient's anatomy registration to his preoperative CT scan data, the navigation system allows zygomatic guiding to its planned reduced location and bone orbital volume restoration control. The feasibility of this technique was checked in 3 patients with major OZMC deformities. Preliminary clinical results are presented.

## MATERIALS AND METHODS

### Preoperative planning

Planning method for primary or secondary surgical treatment of OZMC was previously widely described [2]. Main steps are reminded.

### 3D models generation

The planning workstation provides from a regular CT scan dataset axial and reconstructed coronal and sagittal views, as well as 3D reconstructed surface soft-tissue and skeletal. Practitioner can interactively browse through the different series, manipulate the 3D models or select anatomical points or Regions Of Interest (ROI).

*Target determination*

Whereas the target location for the fractured bone segment is defined symmetrically to the healthy side, a theoretical symmetric skull is computed around a midsagittal plane (fig. 1). This plane is defined by 3 included and manually selected anatomical landmarks.

To take into account inaccuracy due to midsagittal plane manual definition, a double registration, first rigid then elastic, of the computed hemiskull to the bone structures surrounding the fractured zygoma is performed (fig. 2). Thanks to this process, an estimation of zygomatic location before injury has been obtained and will be able to be used as the target location for zygoma repositioning in the following steps of the planning.

*Zygoma segmentation*

Displaced bone is semi-automatically segmented using a spherical ROI which is defined by 4 points belonging to its limit and rather corresponding to the fractured buttresses. This sphere includes the body of zygomatic bone which is the most significant skeletal part for facial symmetry restoration.

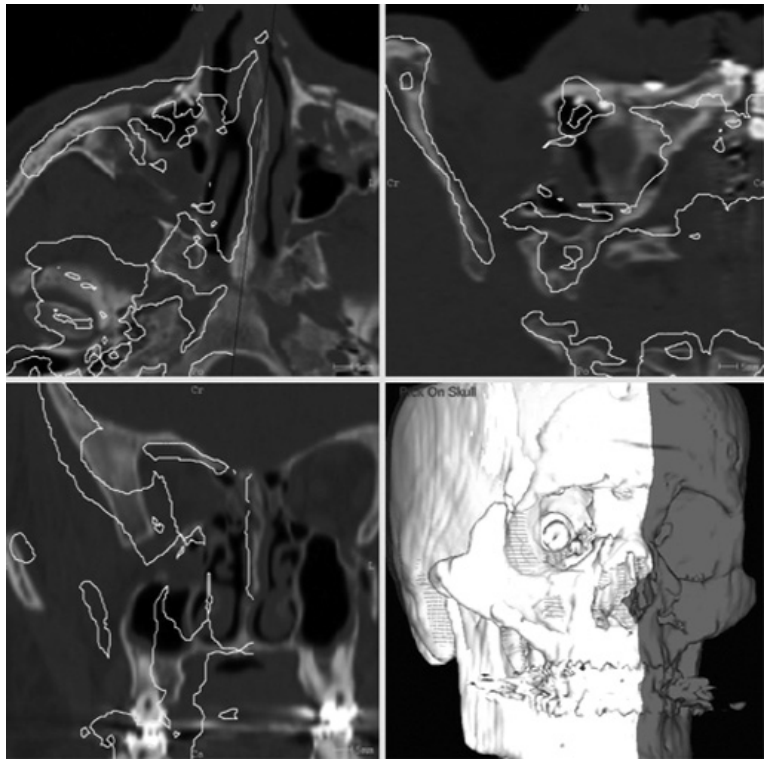

*Fig. 1 : Native skull overimpressed with symmetrical skull computed from the controlateral healthy side around midsagittal plane.*

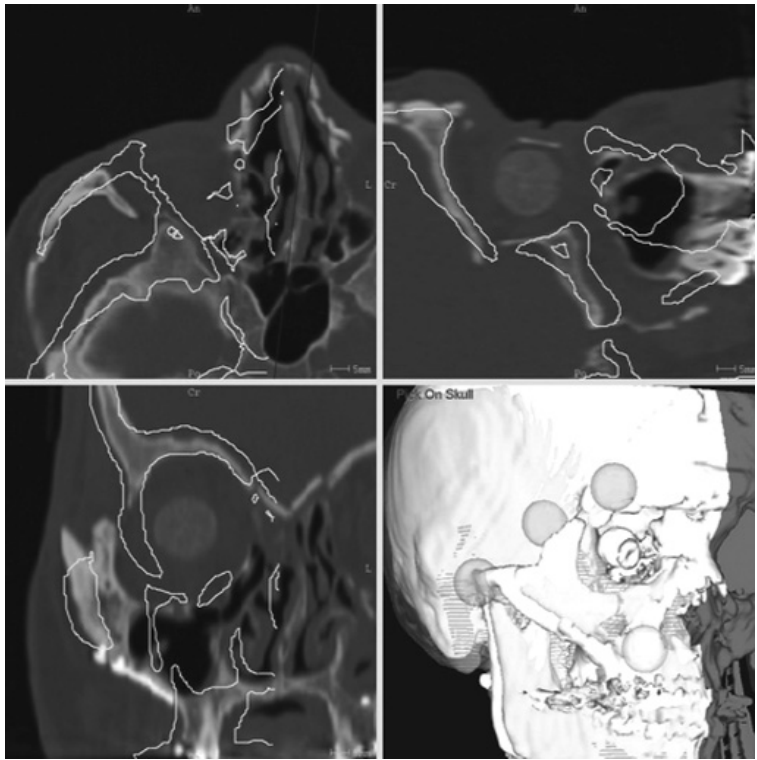

*Fig. 2 : Result from the double registration of symmetrical skull to native skull according to selected surrounding ROI (spheres).*

*Surgical planning for zygoma reduction*

Finally, an automatic rigid registration of the previously segmented fractured bone to the target skull computed in step 2 is performed to determine the best zygomatic bone repositioning to achieve skeletal symmetry (fig. 3). Since matching provides all the better result that initial position of both surfaces is close, a prior point-to-point registration can be required in case of large displacement.

*Non-injured orbit segmentation*

The same as for zygomatic reconstruction planning, orbital boundaries restoration is planned symmetrically from the healthy side. Orbital volume is semi-automatically segmented with a cone-shaped growing ROI defined by its main axis corresponding to the neuro-optic axis and a closing anterior plane (fig. 4).

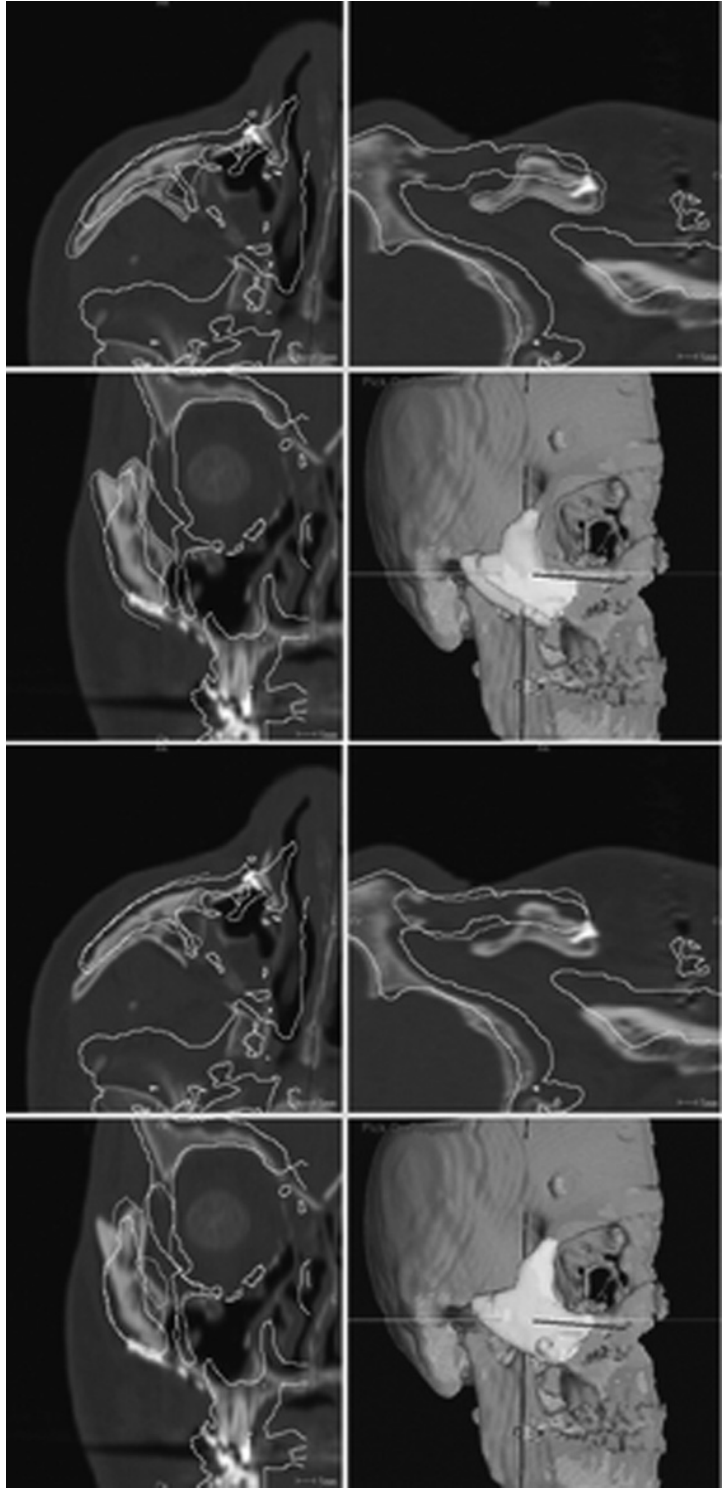

*Fig. 3 : Segmented injured zygoma before (top) and after (bottom) registration to target skull.*

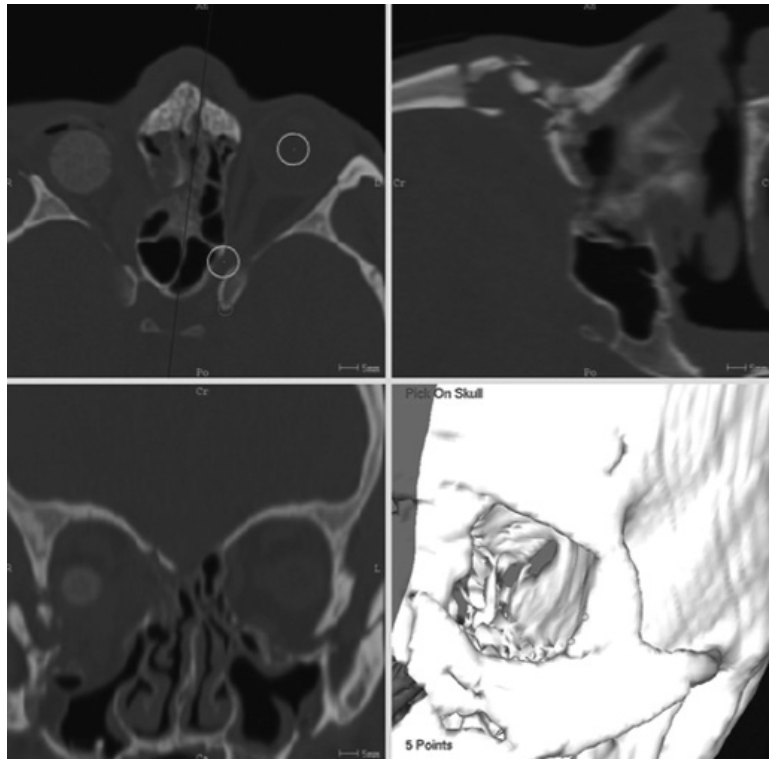

*Fig. 4 : Segmentation of the healthy orbit with a cone-shaped growing ROI defined by the neuro-optic axis and an average closing anterior plane.*

### *Surgical planning for orbital boundaries reconstruction*

A virtual orbit is then computed from the previously segmented orbit symmetrically around the midsagittal plane. Its boundaries are superimposed on the different axial, coronal and sagittal series and will be used as a guiding for accurate orbital walls reconstruction (fig. 5).

### *Intraoperative navigation and guiding*

The Surgetics™ navigation system and the Craniologics™ software (Praxim-Medivision, La Tronche, France) were fitted and used for intraoperative navigation. This navigation system uses a passive and infrared light based optical technology. Novelty of the Craniologics™ software is a double point-surface registration of patient's anatomy to preoperatively acquired CT scan data set, first cutaneous, second skeletal after surgical exposure to improve navigation accuracy around the ROI. Then, location of calibrated surgical instruments in relation to its actual location in the surgical field can be presented in the different series or 3D models and in real time updated.

Software adaptation consists of an additional zygomatic navigation step. First, a Rigid Body is rigidly fixated to zygomatic body by a 2 pins transcutaneous fixation (Digital Zygomatic Frame, DZF) (fig. 6). Next, a point-to-surface registration between zygoma and its model transferred with the planning data set is performed. Thus, zygoma can be guided to the planned best repositioning. Relative position between actual zygomatic location and target location is in real time updated and presented by a distance map between both surfaces and by boundaries of both elements superimposition on axial, sagittal and coronal series (fig. 7).

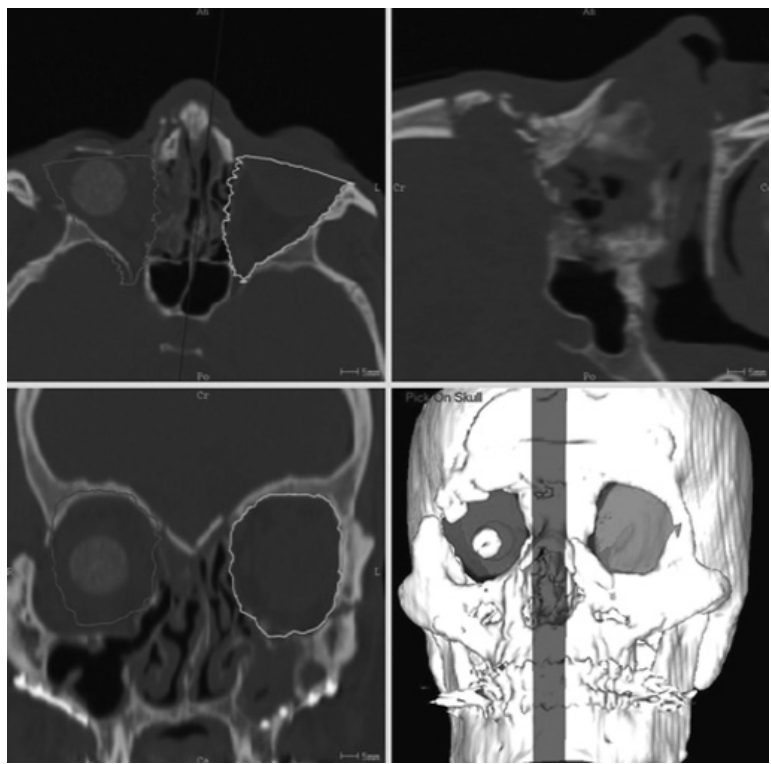

*Fig. 5 : Planning for orbital reconstruction defined by symmetrical orbit from the healthy one around midsagittal plane.*

Navigation interface also provides overimpression of planned orbital reconstructed boundaries on axial, coronal and sagittal series. Accuracy of orbital walls reconstruction or implant placement according to planning is checked by scanning reconstructed orbital walls surface with the calibrated pointer.

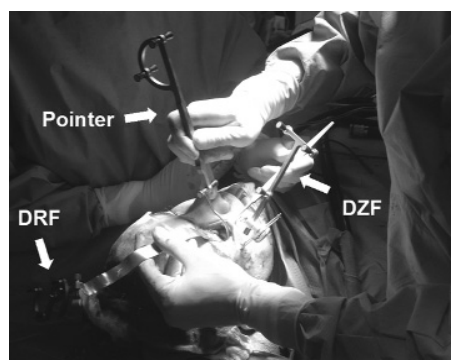

*Fig. 6 : 3DNS setting. DRF : Digital Reference Frame for registering patient's anatomy to its CT scan data set. DZF : Digital Zygomatic Frame for navigating bone fragment.*

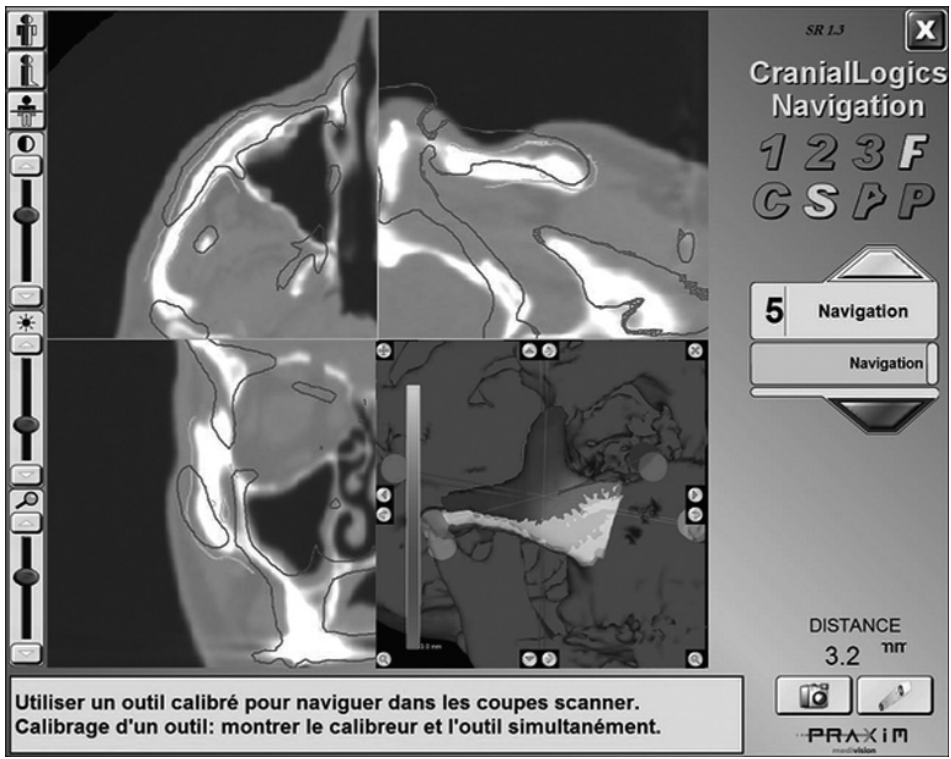

*Fig. 7 : 3DNS interface. Navigated zygoma location is in real time updated and tracked through its boundaries overimpression on axial, coronal and sagittal views in comparison to target location or a distance map between actual and target locations.*

Problem raised by our method was to know if the point-to-surface zygomatic registration according to the small bone surface exposed through the limited surgical approach is enough accurate for Zygoma repositioning. This has been evaluated on a one piece dry skull previously CT scan acquired. 3DNS was set up as previously presented. Mean and max distances of zygomatic surface in DZF referential to the same surface in DRF referential was computed in 10 successive manipulations.

*Patients*

A pilot study of a non-comparative, non-randomized series of 3 patients (table 1) was carried out to check method feasibility. Advisory Comity for People Protection Consenting to Clinical Research approval was obtained through the Toulouse University Hospital, France. Informed consent forms were provided to patients, signed by both sides and collected by investigators.

2 patients had post traumatic malar asymmetry with associated enophthalmos. One of them *(patient 1)* could not have received primary surgical treatment because of initial neurosurgical injury and the second *(patient 2)* was first operated 6 months before in another center. Patient 2 underwent bilateral OZMC fracture and left side was considered as the reference side for controlateral symmetrization. Patient 1 initially lost vision by optic nerve injury and patient 2 by eyeball dilaceration which required prosthetic rehabilitation. The last patient *(patient 3)* had an acute largely displaced zygomatic fracture with severe comminutive fracture of zygomatico-maxillary buttress and orbital rim. No ophthalmologic complication was associated.

Surgical approaches and number of miniplates for zygomatic fixation were individualized according to the patient's fracture. Intraoral upper sulcus and transconjunctival incisions were used in all patients. In secondary treatment, a temporal or coronal approach previously used in a neurosurgical treatment was associated to control zygomatic arch.

# RESULTS

## *Accuracy of the point-to-surface zygomatic registration*

| Determination | Mean (mm) | Max (mm) |
|---|---|---|
| 1 | 0,41 | 0,93 |
| 2 | 0,45 | 0,97 |
| 3 | 0,32 | 0,90 |
| 4 | 0,46 | 0,89 |
| 5 | 0,61 | 1,11 |
| 6 | 0,28 | 0,71 |
| 7 | 0,48 | 0,87 |
| 8 | 0,43 | 0,99 |
| 9 | 0,52 | 1,00 |
| 10 | 0,41 | 0,68 |
| **Mean** | 0,44 | 0,91 |

*Table 1 : Distances computed of zygomatic surface registered in DZF referential to the same surface registered in DRF referential.*

The computed mean distance between the same zygomatic registered in DRF referential and DZF referential was 0,44 mm in mean and 0,91mm in maximum (Table 1).

## *Pilot study*

Computer aided planning and navigation procedures were successfully carried out for all 3 clinical cases. Surgical procedures and postoperative courses were uneventful.

Preoperative planning was completed in 7 to 10 minutes and was judged quite convenient by practitioners. Virtual repositioning alone was useful to conceptualize 3D displacement of zygomatic bone. Navigation and guiding device and specially DZF did not stop performing surgical treatment. Surgical time increase due to 3DNS installation was about 20 minutes. Whereas no anatomical landmarks allowed intraoperatively to check symmetry achievement, zygoma repositioning was done as indicated by intraoperative guiding whatever practitioner thought.

Surgical outcomes are presented (fig. 8, 9, 10). They have been considered as satisfying by both patient and surgeon in all 3 cases. Patient 2 who had initially bilateral OZMC fracture has not been back on his pretraumatic appearance but on a slightly enlarged facial width. However, he was totally satisfied with facial symmetrization. None of the 3 patients complained of the cheek scare.

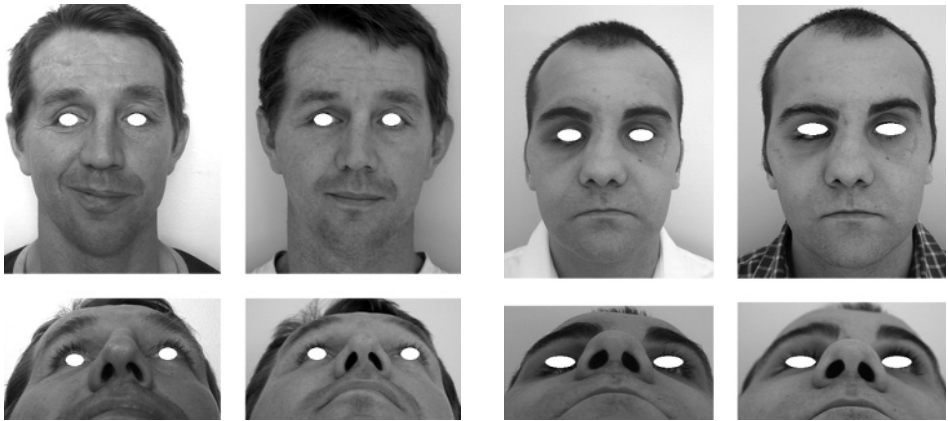

*Fig. 8 : Patient 1. The postoperative pictures (right) in comparison to preoperative pictures (left) show restoration of antero-posterior projection of the zygomatic body and the facial width.*

*Fig. 9 : Patient 2. The postoperative pictures (right) in comparison to preoperative pictures (left) show restoration of antero-posterior projection of the zygomatic body and the facial width. Right prosthetic eyeball remains postoperatively retrusive despite orbital volume reconstruction.*

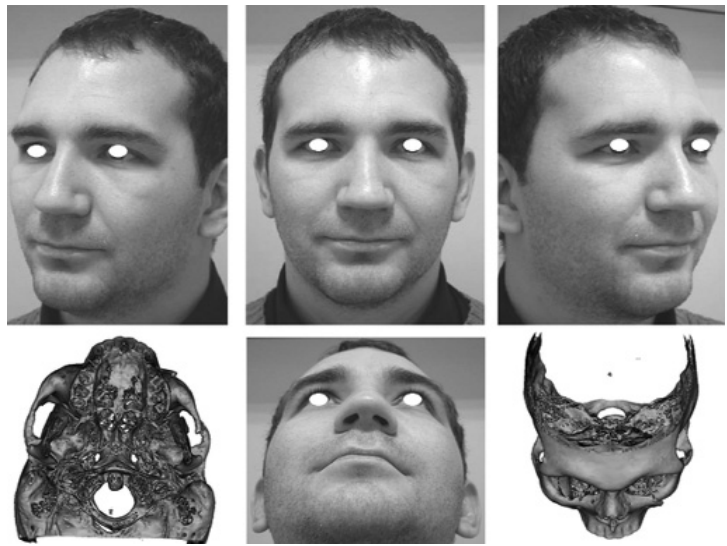

*Fig. 10 : Patient 3. The postoperative pictures (top and centre bottom) show facial symmetry restoration whereas the preoperative CT scan (bottom right and left) points out zygomatic displacement.*

## DISCUSSION

Development of augmented reality techniques seems to be able to help surgical treatment of late or acute OZMC deformities resulting from displaced or comminuted fractures. Preoperative, 3D multiplanar or surfacic reconstruction from CT scan data set is helpful in evaluating the extent and the way of post traumatic deformities. Then, 3D planning system based on 3D virtual models allows segmentation of the skeletal region of interest and interactive repositioning in order to assess different alternatives and retain the best to achieve skeletal symmetry. Finally, surgical procedure can be carried out as planned preoperatively using 3DNS control and guidance. Thus, several suggestions were made addressing whether zygomatic repositioning or orbital reconstruction, planning or operative assistance for surgical treatment [2, 3, 4].

The method we present includes all steps of a comprehensive computer aided protocol from diagnostic to intraoperative assistance, for both zygomatic and orbital reconstruction in primary or secondary treatment. Diagnostic, planning and simulation were automated as much as possible to shorten allocated time and get the process friendly. All 3D skeletal and soft tissue models, fractured zygoma or non-injured orbit are semi-automatically segmented. Zygoma repositioning is automatically planned as the registration of segmented injured zygoma to the target skull. The corresponding mathematical transformation is equivalent to the repositioning movement to apply. So far, manual shift with 6 degrees of freedom in a virtual environment was suggested [4] but seemed time consuming, difficult to achieve in the same time reduction of all 4 zygomatic buttresses and inaccurate in case of dislocated fracture which does not allow visual control of buttresses alignment. Targets for both zygomatic and orbital reconstruction are defined symmetrically from the healthy side around the midsagittal plane. Landmarks-based definition of the facial midsagittal plane without any further correction can give rise to a shift in target skull definition. So, in correction, we suggest to apply a double registration first rigid second elastic between symmetrical hemiskull computed from the healthy side and regions of interest of the native skull surrounding OZMC.

The planned procedure components including 3D native skull, target skull, target orbit and zygoma models as well as mathematical transformation are directly transferred to the operating theatre. OZMC reconstruction outlines can be check similar to the planned ones scanning bone surface with the calibrated pointer, but zygomatic tracking makes proper repositioning easier and shorter. So far, only surface scanning control was performed [4] what means zygomatic osteosynthesis has to be taken down if repositioning is not checked correct.

Primary or secondary OZMC reconstruction was addressed by others workgroups, because it synthesizes requirements and difficulties of computer-assisted preoperative planning and intraoperative assistance in cranio-maxillofacial plastic and reconstructive surgery. However, none of them dealt with orbital and zygomatic reconstruction in the same time while they are anatomically and physiopathologically closely linked. Since no pretraumatic imaging data is available, all groups are agreed to plane reconstruction symmetrically from the healthy side around a midsagittal plane. However, to avoid inaccuracy due to landmarks based definition of the midsagittal plane, either an automatic extraction based on matching of homologous surface areas using an iterative closest point optimization can be used [5] or, as we suggest, a correction by matching the theoretical symmetrical skull to the native skull around the region of interest. Planning of Zygoma repositioning to the target skull was mainly manual what reserves the method for moderately displaced zygomatic fracture without buttress dislocation whereas large displacement, dislocated fracture and secondary treatment are probably the best indications for computed-assisted OZMC reconstruction. Some groups ruled out direct bone guiding fixing DRF onto the segmented bone which should be moved, to use as intra operative assistance the only surface scanning with a calibrated pointer to check reconstruction outline accurate [3]. In our opinion, in selected reconstructive problems, direct bone guiding can be useful to achieve more easily repositioning as preoperatively planned.

This pilot study on the computer-assisted management of post traumatic OZMC deformities show virtual planning for orbital boundaries reconstruction or zygomatic repositioning, as well as intraoperative guiding and verification of the surgical procedure in accordance with the planned one can be successfully carried out. First results seem promising specially in difficult case of late deformities.